
\input harvmac

\def\Title#1#2{\rightline{#1}\ifx\answ\bigans\nopagenumbers\pageno0
\vskip0.5in
\else\pageno1\vskip.5in\fi \centerline{\titlefont #2}\vskip .3in}

\font\caps=cmcsc10

\noblackbox
\parskip=1.5mm


\def\npb#1#2#3{{\it Nucl. Phys.} {\bf B#1} (#2) #3 }
\def\plb#1#2#3{{\it Phys. Lett.} {\bf B#1} (#2) #3 }
\def\prd#1#2#3{{\it Phys. Rev. } {\bf D#1} (#2) #3 }
\def\prl#1#2#3{{\it Phys. Rev. Lett.} {\bf #1} (#2) #3 }

\def\jmp#1#2#3{{\it J. Math. Phys.} {\bf #1} (#2) #3 }

\def\bb#1{{\tt hep-th/#1}}


\def\ket{\rangle}
\def\bra{\langle}

\def\sigmab{\overline\sigma}
\def\psib{\overline\psi}
\def\lambdab{\overline\lambda}
\def\tpsib{\overline{\tilde \psi}}
\def\tpsi{\tilde \psi}
\def\tphi{\tilde \phi}
\def\tD{\tilde D}
\def\epsilonb{\overline \epsilon}


\def\CL{{\cal L}}


\lref\rwit{E. Witten, \plb {86}{1979}{283.}}
\lref\rwitvafa{C. Vafa and E. Witten, \npb {432}{1994}{3.}}
\lref\rseiwitint{N. Seiberg and E. Witten, \npb {426}{1994}{19. \semi}
K. Intriligator and N. Seiberg, \npb {431}{1994}{551.}}
\lref\rseibint{For a review and references see K. Intrilligator
and N. Seiberg, RU-95-48, \bb{ 9509066.} }
\lref\rseidos{N. Seiberg, \npb {435}{1995}{129.}}
\lref\rmonol{C. Montonen and D. Olive, \plb {72}{1977}{117.}  }
\lref\rdimred{J. Harvey, G. Moore and A. Strominger, {\it ``Reducing
S-duality to T-duality"} , EFI-95-01, YCTP-P2-95, \bb{9501022. \semi}
M. Bershadsky, A. Johansen, V. Sadov and C. Vafa {\it ``Topological
reduction of 4-d SYM to 2-d sigma models"}, Harvard preprint,
HUTP-95-A004, \bb{9501096.}}
\lref\rseibintso{ K. Intrilligator and
 N. Seiberg, \npb {444}{1995}{125.}}
\lref\rwitol{ D. Olive and E. Witten, \plb {78}{1978}{97.} }
\lref\rthp {G. 't Hooft, \npb {79}{1974}{276. \semi}
A. M. Polyakov, {\it JETP. Lett.} {\bf 20} (1974) 194. }
\lref\rakmrv{Wess and Bagger, {\it `` Introduction to
Supersymmetry"}\semi  Princeton University Press 1992 \semi
 D. Amati, K. Konishi, Y. Meurice, G.C.
Rossi, G. Veneziano,
{\it  Physics Reports} {\bf 162}, no. 4 (1988) 169.}
\lref\rbogps{ E.B. Bogomol'nyi, {\it Sov. J. Nucl. Phys.} {\bf 24}
(1976) 861. \semi
M.K. Prasad and C.M. Sommerfield, \prd {15}{1977}{554.} }
\lref\rcol{ S. Coleman, Les Houches lectures  Session XXXVII, 1981
  \semi
A. Brandt and F. Neri, \npb {161}{1979}{253.}}
\lref\rgno{ P. Goddard, J. Nuyts and D. Olive, \npb {125}{1977}{1} }
\lref\rgo{ P. Goddard and D. Olive, \npb {191}{1981}{511 \semi}
P. Goddard and D. Olive, \npb {191} {1981} {528.} }
\lref\rhumph{ J. E. Humphreys, {\it ``Introduction to  Lie algebras and
Representation theory"}, Springer Verlag 1983.}
\lref\rkut{ D. Kutasov,  \plb {354}{1995}{315}
\bb{9503086.}}
\lref\rjackreb{ R. Jackiw and C. Rebbi, \prd {13}{1976}{3398.}}
\lref\rsing{ J. Arafune, P G O Freund, C. J. Goebel, \jmp {16}{1975}
{433. \semi}
E. Tomboulis and G. Woo, \npb {107}{1976} {221.}}
\lref\rads{I. Affleck, M. Dine and N. Seiberg, \npb {241}{1984}{493 \semi}
 \npb  {256}{1985}{557.}}
\lref\rosbo{H. Osborn, \plb {83}{1979}{321.}}
\lref\rbw{ F. A. Bais and H. A. Weldon, \prl  {41} {1978}{601.}}
\lref\rswii{ N. Seiberg and E. Witten, \npb {431} {1994}  {484}
, \bb{9408099}}
\lref\rgaunt{J. Gauntlett, \npb { 411} {1994}  {443.} }
\lref\rgh{ J. Gauntlett and J. Harvey,
CALT-68-2017, EFI-95-56, \bb{9508156} }
\lref\rblum{ J. Blum, \plb  {333} {1994} {92.} }
\lref\rssz{S. Sethi, M. Stern, E. Zaslow,
HUTP-95/A031, DUK-M-95-12, \bb{ 9508117.} }
\lref\rcfns{M. Cederwall, G. Ferretti, B. Nilsson, P. Salomonson,
Goteborg-ITP-95-16, \bb{9508124.} }
\lref\rmo{ C. Montonen and D. Olive,  \plb {72} {1977} {117.} }
\lref\rswi{ N. Seiberg and E. Witten, \npb {426} {1994} {19},
             \bb{ 9407087} }
\lref\rsen{ A. Sen, \plb {329} {1994} {217}, \bb{ 9402002.} }
\lref\rengwi{ F. Englert and P. Windey, \prd {14} {1976} {2728}  }
\lref\rfulha{ W. Fulton and J. Harris, {\it ``Representation theory"},
 Springer Verlag 1991. }
\lref\rlestr{ R. Leigh and M. Strassler,
\npb {447} {1995} {95}, \bb{9503121.}  }
\lref\rgp{D.J. Gross, R. Pisarski and  L. Yaffe, {\it Rev. Mod. Phys.}
{\bf 53} (1981) 43.   }


\line{\hfill PUPT-95-1580}
\line{\hfill {\tt hep-th/9512063}}
\vskip 1cm

\Title{\vbox{\baselineskip 12pt\hbox{}
 }}
{\vbox {  \centerline{ Some properties of unstable}
           \centerline{monopoles in Super-QCD }
\medskip
\centerline{ }  }}

\centerline{$\quad$ {\caps J. L. F. Barb\'on  and S. Ramgoolam }}
\smallskip
\centerline{{\sl Joseph Henry Laboratories}}
\centerline{{\sl Princeton University}}
\centerline{{\sl Princeton, NJ 08544, U.S.A.}}
\centerline{{\tt barbon@puhep1.princeton.edu}}
\centerline{{\tt ramgoola@puhep1.princeton.edu }}
\vskip 0.4in

We study embeddings of the Prasad-Sommerfield monopole solution in
$SU(N_c)$ Super-QCD  $(N_c \ge  3)$, where the  role
 of the Higgs field is played by the
squarks in the fundamental representation. Classically,
the resulting
configurations  live in a phase with unbroken
$SU(k)$ subgroups of $SU(N_c)$ ( as a result they are not topologically
stable).
The structure of zero modes of these monopoles is such that they
 can be naturally interpreted as massive chiral
superfields, with
R charge one and baryon number zero. They transform
in the adjoint representation of a dual gauge group
defined using the Goddard-Nuyts-Olive (GNO) framework.
We  discuss the possible
applications of these monopoles to $N=1$ duality, and more generally
 the possibility of relating GNO type dual gauge groups to those
appearing in N=1 duality.


\Date{11/95}


\newsec{Introduction}
Considerable evidence has been found for a version
 of electric-magnetic duality that  governs
 the low energy dynamics of certain $N=1$ supersymmetric
theories \refs\rseidos , \refs\rseibint.
Among the more striking aspects is the statement that
in the infrared  $SU(N_c)$ theory with $N_f$ flavours
( $N_c + 2 \le N_f \le 3N_c$ ) is an interacting conformal
field theory which has a dual description in terms of magnetic
variables
with gauge group $SU(N_f-N_c)$ containing $N_f$ quarks and $N_f^2$
mesons.
This  differs from previous electric-magnetic duality
conjectures (mainly in the context of extended $N=2$ and $N=4$ SUSY
theories) in various aspects.
Since  the low energy theory has a manifest
non-abelian gauge symmetry,  the simple abelian duality
transformations of Maxwell theory cannot be used, even at a formal
level, to understand the operator mapping.
In the $N=2$ and  $N=4$ dualities \refs{ \rswi, \rswii ,\rsen }
 monopoles \refs\rthp\rbogps ,
 play a fundamental role,
being the elementary degrees of freedom in the dual formulations.
In these theories the relevant monopoles
saturate a BPS bound and  live in small representations
\refs\rwitol , so although the semiclassical construction of states
starting from non-trivial solutions is only valid, in the
 region of large Higgs vevs, these states can be argued  to survive
in the strong coupling region. In  $N=1$ theories in $4D$,  massive
states cannot be protected by such a BPS saturation, so there is no
guarantee that states found by semiclassical treatments will survive
in the strong coupling region.
In these respects $N=1$ duality is perhaps more mysterious
than other field theoretic dualities proposed so far.
  For some special $N=1$ theories, related
to theories with higher supersymmetry or having an abelian Coulomb phase,
duality can be related to monopole physics \refs\rseibintso.
One may naturally
 ask to what extent these relations survive in
the $SU(N_c)$ examples.
The first problem one meets is that  super-QCD does not seem to support
standard monopole configurations of 't Hooft-Polyakov type,
which are constructed using adjoint Higgs fields,
 so there
are not even natural candidates in the semiclassical region.

In this
paper we describe a class of non-singular finite energy
 monopole solutions of such theories.
Like any classical monopoles living in a phase with
unbroken $SU(k)$ groups they are not
 topologically stable.  They have an interesting zero mode structure
and  admit an interpretation as an  adjoint chiral superfield.
The interpretation of the gauge quantum numbers is done using  the
framework of Goddard, Nuyts and Olive \refs\rgno\
(see also \refs\rengwi),   with the
additional feature that we discuss and use
 the freedom  of associating a dual
group to a  sublattice of the lattice of magnetic weights
allowed by Dirac quantization. We comment more generally
on the possible applications of this freedom in $N=1$ duality.
 Although it  is conceivable
 that no vestige of $N=1$ duality in the $SU(N)$  models
 can be captured with
semiclassical methods, we  speculate on
possible scenarios where these monopole solutions
would be related to duality.

\newsec{Construction of Classical  Solutions}

  We  construct a class of monopole solutions of $SU(N_c)$
 ($N_c > 3$ ) Yang Mills theories coupled to fundamental matter.
We do this by making an ansatz with non zero fields living only in an
  $SO(3)$ subgroup, and then reducing the equations to those of an
$SO(3)$ theory coupled to an adjoint of the group.
 Consider the model Lagrangian:
$$
L = -  {1\over 2} tr F_{\mu \nu}F^{\mu \nu} - (D_{\mu}\phi)^{\dagger }
          (D^{\mu}\phi)$$
For simplicity we start with $SU(3)$ gauge group and take $\phi$ as
a complex scalar transforming in the fundamental representation.
  Let $H$
be the real $SO(3)$ subgroup. The
fundamental  of $SU(3)$ transforms as the adjoint of $SO(3)$,
whose Lie algebra we denote by $h$.
We now consider
 an ansatz with $A_{\mu}^a=0$ for $a \in \bar h$, the complement
of $h$, and $\phi$ real. With such an ansatz the field strength is also
zero outside $h$, and the $SU(3)$ covariant derivative
reduces to the $SO(3)$ covariant derivative. The field equations
are:
\eqn\eom{\eqalign{
&(D^{\mu}  F_{\mu \nu} )^a = i  \phi^\dagger T^a D_{\nu} \phi - i
( D_{\nu}\phi)^\dagger T^a \phi \cr
&(D^2 \phi)^a = 0 \cr}}
The scalar field equations and the gauge field equations for
$a \in H$ reduce trivially to the corresponding equations
for the $H$ gauge theory. It remains to check that the
gauge field equations for $a \in \bar h $ are also satisfied. This
amounts to showing that, for $a \in \bar h $ :
\eqn\ccel{
0 = i\phi^t T^a \partial_{\nu} \phi - i( \partial_{\nu} \phi )^t T^a \phi
   + \sum_{ b \in h} A_{\nu}^b \phi^t ( T^aT^b + T^bT^a) \phi .}
We have used reality of $\phi$ to rewrite $\phi^\dagger$ as
$\phi^t$, the transpose.  The first two terms cancel because, for
$a\in \bar h$, $T^a$ is symmetric. To see the cancellation
of the remaining two terms, use the fact that
$$ \{ T^a,T^b \} = k \delta^{ab} + d^{abc} T^c .$$ The $d^{abc}$ are real
because the $T$'s are hermitian. For $a\in \bar h$, and $b \in h$,
$T^c$ is imaginary, and therefore antisymmetric.
This guarantees that the
last term is zero.

Having reduced the $SU(3)$ equations to $SO(3)$ equations we
know that the Prasad-Sommerfield solutions solve them. So there
are finite energy,  non-singular
 monopole, dyon, and multimonopole solutions to the
equations of $SU(3)$ gauge theory coupled to fundamental matter.
 These  solutions
    can be  embedded in $SU(N)$ coupled to
scalars  in the  fundamental representation.
  This is done by using an ansatz for $A_{\mu}^a T^a$
   where the only non-zero $A_{\mu}^a$ correspond to generators
   of an $SU(3)$ subgroup associated with say the
   first $3 \times 3$ block of the $N\times N$ matrices
   of the fundamental of $SU(N)$. And the only non-zero scalars
    are taken to live in the first three entries of the $N$ dimensional
    vector.
 The reduction of equations of $SU(3)$ to those of $SO(3)$
   used the reality of the subgroup, and  a similar reduction works for
  $SU(N)$ and $SO(N)$. Note that the usual embeddings of the
  BPS solution when the matter is in the adjoint of the original
  gauge group  do not make use of such reality conditions.

 The scalar field expectation values
at spatial infinity
 break the gauge symmetry from $SU(N_c)$ to $SU(N_c-1)$ for the
simplest solutions. Since $\pi_2 ( SU(N_c)/SU(N_c-1) )
 \sim \pi_1 (SU(N_c-1)$ is trivial, these solutions are not
topologically stable. It has been shown that configurations with the
long distance magnetic fields of such monopoles are actually also
locally unstable to perturbations of the gauge
 fields in the unbroken $SU(N_c-1)$ subgroup \refs\rcol  .

 We now describe how to embed these solutions in $N=1$ supersymmetric
  gauge theory. Following the conventions of
 \refs\rakmrv ,
 the bosonic terms are:
 \eqn\lag{\eqalign{
L = -  {1\over 2}tr F_{\mu \nu}F^{\mu \nu} - (D_{\mu}\phi)^ {\dagger }
          (D_{\mu}\phi) - (\tilde D^{\mu} \tilde \phi)
      (\tilde D_{\mu}\tilde \phi)^{\dagger} - { e^2 \over 2}
 [ \phi^{\dagger}T^a\phi - \tilde \phi T^a \tilde \phi^{\dagger} ]^2
    }}
Where $\phi$ and $\tilde\phi$ denote the scalar components of the
quark and antiquark superfields respectively. Also, ${\tilde D}_\mu$
is the covariant derivative in the complex conjugate fundamental
representation of $SU(N)$.
Choosing an ansatz with $\phi =\tilde \phi^\dagger $ and  real,
the $D$-term clearly vanishes,
the purely bosonic equations reduce to those of the
  model Lagrangian considered
above,  except that we now have two scalar fields.
In this way we can carry over the solutions
from the model Lagrangian by a simple rescaling. The corresponding
Bogomolnyi equations take the form
\eqn\bogo {
B_i =\pm 4 D_i \phi =\pm 4({\tilde D}_i {\tilde \phi})^{\dagger}
}
In writing these equation we made use of the identification between
the vector of $SO(3)$  and the fundamental representation of $SU(3)$.

Conditions for the vanishing of $D$-terms for arbitary $N_f$ and $N_c$
have been investigated in \refs\rads\
and yield the well known moduli space of vacua which has been studied
classically and quantum mechanically \refs\rseibint.
The simplest solution described above satisfies
boundary conditions appropriate for the meson field
matrix, $M^i_{\tilde i}$
  acquiring an expectation value of rank one, the case in which we
will concentrate in the rest of the paper. However, the construction
of solutions can be easily generalised to cases where $M$ has rank smaller
than $N_c/3$ by using different   $SO(3)$ subgroups
of $SU(N_c)$ which couple to different
flavours. Note that, at the level of the classical Lagrangian
these solutions exist for any $N_f$. After taking into account
quantum corrections \rseidos , these configurations do not have finite
energy unless $N_f \ge N_c+2$ where the classical moduli space is the
same as the quantum moduli space.

These solutions  break the $N=1$
supersymmetry. This is clearly seen, for example, from the gluino
transformation rules
  \eqn\ssy{ \delta_{\xi} \lambda^a = i \xi D^a + \sigma^{\mu\nu}\xi
 F^a_{\mu\nu}. }
due to the non vanishing field strength (the $D$ term was arranged to
be zero).
The resulting fermionic zero modes in the monopole background
 lead to states filling out a representation of supersymmetry, as
discussed in detail later.
In practice, the absence of unbroken supersymmetries means that we do
not have much control on the strong coupling physics of these
solutions.

\newsec{ Embeddings and gauge quantum numbers}

  The gauge quantum numbers of the monopole can be described
in the framework developed by Goddard, Nuyts and Olive (
 GNO)  which has been useful in
investigations of Montonen-Olive  duality \refs\rmo\ in $N=4$ theories
 \refs\rdimred . In this section we briefly review their construction,
discuss the possibilities of some simple variations
on their definition of dual group, while keeping the key property that
the weights of the dual group are magnetic weights ( see definition below)
 of monopoles. Using these remarks we deduce that the monopoles
constructed above transform as adjoints. We will  also discuss
in some generality the applicability
of these definitions to theories with less than $N=4$ supersymmetry.

 The long distance magnetic field
defines an element $M$ of the Lie algebra  $h$ of the
unbroken gauge group:
 $$F_{i j } = \epsilon_{i j k } {x_k \over { x^3}} M .$$
By a gauge transformation, $L=eM$ can be taken to lie in the Cartan
subalgebra $h$. Using the metric this determines an element $\tilde L$,
 called a magnetic  weight,
of the dual vector space $h^*$. $L$ and $\tilde L$
are related by $ < \tilde L, h_1 >  = ( L,h_1)$, for any
$h_1 \in h$, where the LHS is the evaluation
of the functional $\tilde L$ on $h_1$ and $ (.,.)$ is the metric on the
Cartan. $(.,.)$ induces a metric on $h^*$ : $$ (\tilde L_1 , \tilde L_2) =
 (L_1, L_2). $$
The Dirac quantization
condition,
\eqn\Dirq{  e^{4\pi i L}=1, }
 in conventions where $D_\mu = \partial_{\mu} +
 ie A_{\mu}^a T^a$,
allows $\tilde L$ to live in a lattice, which we will call the
Dirac lattice.
GNO found that this lattice coincides with the weight lattice of
a dual group $G^{v}$.
The roots of $G^{ v}$ are proportional to the roots of
$G$ :  $\alpha^{ v} =
{ \alpha \over { (\alpha , \alpha)}} $,  and  the global structure
of $G^{v}$ is determined by
\eqn\glob{ G^{ v} = \widetilde G^{ v} / k(G^{ v}) }
where $  k(G^{ v})$ is the quotient of the weight lattice
of $G$ by the root lattice of $G$, and ${\widetilde G}^v$ denotes the
universal cover of $G^v$.  Since $G$ is a subgroup of the
broken
gauge group, $\tilde L$ can also be related to an element of
the dual of the broken group \refs\rgo. In the following discussion
we will characterize the monopole in terms of  the GNO dual of the
  unbroken
group, but the discussion can be extended to  the broken group.
We note that the set of
$\alpha^{v}= {k \alpha \over { (\alpha , \alpha)}}$ defines
a root system satisfying the usual axioms  for root systems \refs\rhumph ,
(hence allows reconstruction
of a dual group ) for any $k$. Note also
that for any $k$, $(\alpha^v )^v = \alpha$.
Dual root systems  defined using different values of $k$  are
isomorphic to each other as root systems
(i.e the isomorphism preserves angles),
and hence determine isomorphic Lie algebras.

It is useful to bear  in mind that
in some cases there may be different ways of defining
$G^v$ such that $\Lambda (G^v)$ is the lattice of magnetic weights.
Consider the case of $SU(2)$ theories. The basic  (spherically symmetric )
BPS monopole
has a magnetic weight ${\alpha\over { (\alpha,\alpha)}}$.
The prescription of GNO defines the dual group by
giving its root as ${\alpha \over{(\alpha,\alpha)}}$, and
its fundamental group as $Z_2$. This means that the dual
group is $SO(3)$ and the BPS monopole transforms as the adjoint.
This is appropriate in $N=4$ where the spins are consistent
with the monopole being dual to a gauge boson. An alternative way to
define a group whose weight lattice is that generated by the basic BPS
monopole, is to use $\alpha^v = 2 { \alpha \over { (\alpha , \alpha)}}$
and declare the fundamental group to be trivial. This picks out
$G^v$ as $SU(2)$ and the spherically symmetric
 BPS monopole transforms as the fundamental
representation of this dual group. This is appropriate in interpreting,
from the GNO point of view, the self duality  of $N=2$   $SU(2)$ theory
with $N_f=4$ described in \refs\rswii.

Another simple variation on the GNO construction we will
consider is to define the dual group whose weight lattice is
a sublattice of the Dirac lattice. We will see that this variation
is necessary in order to have a simple assignment of gauge quantum
numbers for the monopoles of section two. It will also
allow us to address
a puzzle which appears if one tries to understand
some aspects of the  $N=1$ duality of \refs\rseidos\ using the GNO
construction.
  A direct understanding of the duality between $SU(N_c)$ and
$SU(N_f-N_c)$ appears hard, so one  might hope to
understand the duality at the
point $N_f=2N_c$  in terms of the GNO type of construction
and relate  the other cases to these self-dual points using
 flows described in \rseidos\ ( deriving the general duality from
the self dual case has been discussed in the context of relations
with $N=2$ in \refs{\rseidos, \rlestr } ). The puzzle is that
even at the self-dual point,
the standard GNO definition gives the dual group of $SU(N_c)$
as $SU(N_c)/Z_{N_c}$ which does not have fundamental representations.
On the other hand both the electric and magnetic theories  in
\rseidos\ have fundamental quarks.
We will show that considering sublattices of the Dirac
lattice suggests a
possible resolution of this puzzle.

   We now characterize the gauge quantum numbers of the
monopoles constructed in section 2
using the formalism of GNO.  For simplicity consider $SU(N_c)$
with $N_f=1$ in a vacuum where $\langle \phi_1 \rangle
 =  \langle \tilde \phi_1 \rangle \neq 0 $. A simple class of embeddings
of $SO(3)$ generalizing  the one discussed for $SU(3)$ allows us to
construct solutions for this vacuum. These are characterized by three
distinct integers $ (1jk)$ picked from $1$ to $N_c$. For each choice
$(1jk)$ the non-zero components of $\phi$ for the solution
written in 't Hooft Polyakov form \refs\rthp, are
$\phi_1$, $\phi_j$ and $\phi_k$. Rotating to a gauge where $\phi$
 points in a fixed direction \refs\rsing ,
we arrange for the only non-zero component to be $\phi_1$. The
long distance field in such a gauge is
$ F_{a b }  =\pm i [ E_{jk} - E_{kj}] \epsilon_{a b c } {x_c \over {e x^3}}
 $, the matrix in the chosen $SO(3)$ which
annihilates $\phi_1$. In this way we can construct $(N_c-1)(N_c-2)$
monopole solutions.

 We conjugate  $\pm [ E_{jk} - E_{kj}]$ by
$e^{ { i\pi\over 4} (E_{jk} + E_{kj}) }$
 to get  the set
of matrices $\pm (H_j - H_k)$
   where
$H_j $ is the matrix with $1$ in the $j$'th diagonal
and $0$ everywhere else. So we have a set of $(N_c-1)(N_c-2)$ magnetic
weights. These are all related by $W(G)$,
the Weyl group of $G$, and are gauge equivalent in $G$.
Now if $\alpha^v = {k \alpha \over {(\alpha,\alpha)}}$,
and $W_{\alpha} (\beta )$ is the Weyl reflection of
$\beta$ in the hyperplane perpendicular to the root $\alpha$,
 $W_{\alpha} (\beta)
= ( W_{\alpha^v} ( \beta^v ) )^v $. So
 the magnetic weights lie  in one orbit of $W( G^{ v})$.
In addition, for each magnetic weight in the set, there is also
the negative of the weight present in the set.
We want to recognise this set of  weights of the GNO dual group,
 as the non-zero weights
of some  representation of $G^{v}$. The behaviour of the set
under   $W( G^{v})$ rules out reps
like the fundamental which are not self-conjugate.

The roots of $SU(N_c-1)$ take the form $L_i-L_j$ ( $i$ and $j$
are distinct integers between $1$ and $N_c-1$), where the $L_i$ are
linear functionals on the Cartan subalgebra of $SU(N_c-1)$ defined by
$< L_i, \sum a_j H_j > = a_{i}$, and
$\sum a_j H_j $ is any traceless diagonal matrix.
Pick a metric on the Lie algebra,  some  arbitrary constant
$c$ times
trace in the fundamental. With this metric, if $L= H_j-H_k$,
we have $\tilde L = c (L_j - L_k)$, since $c ~ tr (L  \sum
a_j H_j ) = < \tilde L, \sum
a_j H_j ) >$.
Now if $\alpha = L_i - L_j$, the corresponding element of
$h$ is $ {1 \over c} (H_i -H_j)$, so
$(\alpha , \alpha) = c ~ tr {( H_i - H_j)^2 \over {c^2} } = 2/{c}$.
Therefore $ {\alpha \over { (\alpha, \alpha)}} = {c \over 2} ( L_i - L_j ) $.
It follows that the magnetic weights $\tilde L$ of the monopole
are twice the  roots of the GNO dual group.
( For comparison, note that
the BPS monopole as a solution to $SU(2)$ gauge theory
 with adjoint Higgs
defines a magnetic weight equal to ${\alpha \over { (\alpha,\alpha)}}$ and
has the minimal Dirac unit of charge according to definition  \Dirq ).
  The smallest representation  containing these weights
is associated with a Young diagram having two columns of length $N_c-2$
and two columns of length one. However such a representation
has other non-zero weights as well, which cannot be extracted from the
monopoles, so assigning the monopoles to such
representations is problematic.

Instead define the roots of the dual group by  $\alpha^{ v} =
{2 \alpha \over {(\alpha, \alpha)}}$. And fix   the
global structure of the group by identifying the
root and weight lattices, so that the dual group is
$SU(N_c-1)/Z_{N_c-1}$. Then the  magnetic weights of the
monopoles we have constructed
are equal to the roots of the dual group, and hence
 correspond to the adjoint of the dual group. With this definition of dual
group, the $(N_c-1)(N_c-2)$ magnetic weights of the monopole
exhaust the set of non-zero weights of the adjoint representation.
We may  physically motivate this definition of the dual group.
 It appears likely
that not all magnetic weights
allowed by Dirac quantization in $SU(N_c)$ theories coupled to fundamental
matter  correspond to non-singular finite energy monopoles,
and the only such
monopoles  might be
 multimonopoles based on those we have constructed ( with magnetic
weights being integer linear combinations  of those we have described).
 If this is indeed true,
then the  dual group defined above  is distinguished
  in that its  weight lattice is the set of magnetic
weights of non-singular, finite energy, classical
  monopoles in super QCD.

The following remark is not directly relevant to the interpretation
of the monopoles we are studying in this paper,
 but is suggested by the
above discussion.
It may be useful in trying to relate monopoles to duality in $N=1$,
to explore the possibility of constructing the dual group,
using  some sublattice of the Dirac lattice, not necessarily the sublattice
we chose above.
If we use the Dirac lattice, the dual group of $SU(N)$ is $SU(N)/Z_N$, which
does not have fundamental representations (whereas in $N=1$
duality both electric and magnetic sides can have
fundamental quarks). However  the weight lattice of
$SU(N)/Z_N$ contains  sublattices isomorphic, by a rescaling,
 to an $SU(N)$ weight lattice.  To see an example,
 note that the weight lattice
of $SU(N)/Z_N$ is generated by $L_i -L_j$
 with the relation $\sum L_i = 0$ ( see for example \refs\rfulha\ ).
In this lattice $NL_1 = (L_1 -L_2) + (L_1-L_3) + \cdots (L_1-L_N)$. The
$NL_i$ generate a sublattice with the relation $\sum NL_i=0$, which is
isomorphic by a rescaling to an
 $SU(N)$ weight lattice. This suggests a definition of the GNO type dual
  group of $G = SU(N)$
by the prescription $\alpha^v = { N \alpha\over{(\alpha ,\alpha )}}$,
and $G^{v}$ simply connected. With this prescription $G^v$ is isomorphic as
a group to $SU(N)$, and its weight lattice is a sublattice of the Dirac
 Lattice. So the dual group defined
using such a  sublattice might be more closely
 related to the $N=1$ dual group
at the self dual points, e.g $N_f=2N_c$ of \refs\rseidos.
Restricting to this  sublattice would have to be motivated by
some extra  physical input in addition to the Dirac condition
(e.g stability).
It is extremely interesting that the  weights of the fundamental
 and antifundamental representations  of the
dual group ( to  $SU(N)$  )
defined in this way are actually the magnetic weights
of  spherically symmetric monopoles constructed as solutions to
$SU(N)$ gauge theory in \refs\rbw . These monopoles are constructed
using the maximal embedding of $SO(3)$ in $SU(N)$,
and live in a phase
where $SU(N)$ is broken to $SU(N-1) \times U(1)$.
 These monopoles require  adjoint matter for
their construction, and
will not be  discussed further in this paper.

\newsec{Zero mode quantum mechanics}

 The bosonic  solution described in the previous section
 can be acted on by the symmetries it breaks to generate  other solutions
of the same energy.
For example when $N_f > 1$ one can start with a solution
where   $M_1^1 $ tends  to a non-zero
constant
at infinity, and rotate by $SU(N_f)$ matrices to get a family
of solutions parametrized by $(SU(N_f)\times SU(N_f) )/
(SU(N_f-1) \times SU(N_f-1)) $.
The monopoles are $SU(N_f-1)$ singlets:
but this is consistent with it being  a fundamental
or a singlet of $SU(N_f) \times SU(N_f)$.
Even if the vacuum order parameter is in the fundamental of $SU(N_f)\times
SU(N_f)$, the monopole can be a singlet.  A familiar example is the $N=4$
theory, where the $O(6)$ symmetry is spontaneously broken, and the
monopole, being dual to the $W$ boson, is interpreted as a singlet of
$O(6)$.
 One can also use
the $U(1)$ symmetries to rotate $\phi$ and $\tilde \phi$ while leaving
their magnitude constant. However, in the monopole solution, these
symmetries are spontaneusly broken by the vacuum boundary conditions
of the scalar fields at spatial infinity. Thus
 these rotations do not give rise
to normalisable zero modes, or excitations of finite charge, so they
will not be quantized in the following discussion of the quantum numbers
of the monopole.

The low energy dynamics of the monopoles described in the previous
section is dominated by their unstable character. At best, they can be
quantized as unstable resonances, possibly with an interpretation
similar to sphalerons. However,  their quantum numbers in a single particle
Hilbert space are still determined by the effective quantum mechanics of the
zero modes around the classical solution.
For similar recent discussions in $N=2$ theories
see \refs{\rgaunt,\rblum,\rswii,\rgh,\rcfns, \rssz}.
In this paper we
 will only consider the single monopole sector, where one finds the
usual collective coordinates associated with space translations and
charge rotations, leading to dyonic excitations of the monopole, as
well as  fermionic zero modes, whose quantization leads to
non trivial spin degrees of freedom. The corresponding fermionic
equations of motion are given by
\eqn\zmeqs
{\eqalign{ \sigmab^{\mu}D_\mu \lambda^a    &= \sqrt{2} e \sum_{r=1}^{N_f} (
\psib_r T^a \phi_r +  \tphi_r T^a \tpsib_r ) \cr
\sigmab^{\mu} D_{\mu} \psi_r &= -\sqrt{2} e \lambdab^a T^a \phi_r \cr
\sigmab^{\mu} \tD_{\mu} \tpsi_r &= -\sqrt{2} e \lambdab^a \tphi_r T^a
}}

Since the PS monopole breaks all the supersymmetries, an obvious
solution of these equations is given by the supersymmetry variation of
the bosonic PS solution\foot{Note that, unlike the flavor symmetries,
the supersymmetry is $not$ spontaneusly broken by the vacuum boundary
conditions at infinity. Supersymmetry is asymptotically restored at
long distances, and it makes sense  to quantize these zero modes as
collective coordinates.}.
We  find a four parameter family of zero modes
\eqn\zmodes
{\eqalign{\lambda_{\epsilon}& = - {i\over \sqrt{2M}}  B^i_{PS}
\sigma_i\, \epsilon \cr
\psi_{\epsilon} &= {i\over\sqrt{M}} D_i \phi_{PS}\, \sigma^i\, \epsilonb
\cr
\tpsi_{\epsilon} &= {i\over\sqrt{M}} \tD_i \tphi_{PS}\, \sigma^i
\,\epsilonb }}
where $\epsilon$ is related to the anticommuting parameter
 $\xi$ of \ssy\  by
 $\epsilon = { \xi \over {\sqrt {2M}}}$.
 These modes solve
\zmeqs\ with non trivial Yukawa terms along the $r=1$ flavour
direction, since $\phi_{PS}= \tphi_{PS}^{\dagger}$ vanish for $r\neq 1$.
The remaining equations for the $r>1$ flavours are
\eqn\others
{\sigmab^{\mu} D_{\mu} \psi_r = \sigmab^{\mu} \tD_{\mu} \tpsi_r =0}
and have no normalizable solutions in the PS monopole background. To
see this, let us assemble $\psi_r$ and $\tpsi_r$ into a Dirac spinor
$\Psi_r$. Then, eq. \others\ is the $G=0$ case of
\eqn\jacreb
{i\gamma^{\mu} D_{\mu} \Psi_r + iG[\phi_{PS}, \Psi_r ] = 0 }
where the commutator is to be regarded as the adjoint action of
$SU(2)$. This is consistent because the PS solution is in the vector
representation of the distinguished $SO(3)$ where the monopole sits,
and this is equivalent to the adjoint action in \jacreb. The
particular case $G=1$ is well known as it represents the zero mode
equation for the fermionic component of a $N=2$ vector multiplet.
Also, if we define $A_0 \sim \phi_{PS}$ we have the zero mode equation
for adjoint fermions in a particular four dimensional instanton
background.

The  analysis of Jackiw and Rebbi \refs\rjackreb\ shows that
\jacreb\ has two chiral normalizable zero modes along each flavour
direction as long as the adjoint Yukawa term is non zero, $G\neq 0$. A
simple inspection of their results reveals that both zero modes become
non normalizable as $G\rightarrow 0$.
For the PS monopole configuration, normalizing the scalar expectation
values $\bra\phi\ket =1$, the solutions have he form
$$
\Psi^a \sim [f_1 (r) {\hat r}^a \sigma_i {\hat r}^i + f_2 (r) (\sigma^a
-{\hat r}^a \sigma_i {\hat r}^i)]\chi
$$
with $\chi$ a Weyl spinor and $a$ the $SU(2)$ adjoint index. The
radial functions $f_1 (r)$ and $f_2 (r)$ must be normalizable in the
norm $\|f_i \| = \int |f_i |^2 r^2 dr$ and are related via
\eqn\rel
{ f_1 (r) = {\rm sinh}r \left( f_2^{'} (r) + {1-G \over r} f_2
(r)\right) + G\,\, {\rm cosh}r \,f_2 (r)
}
In terms of the funcion $u(r)$ defined as
\eqn\udef
{f_2 (r) = r^{G-3 \over 2} ({\rm sinh}r)^{-{1-G \over 2}} \,\, u(r) }
the zero mode equation reduces to the following zero energy
Schr\"odinger problem:
\eqn\schr
{ -u{''} (r) + V_{\rm eff} (r) u = 0 }
The effective potential $V_{\rm eff}$ is a monotonic decreasing
function with asymptotics $V(r\rightarrow 0) \sim 2/r^2 $ and
$V(r\rightarrow +\infty) \sim (1-G)^2 /4  + O(1/r)$.
There is one regular solution at the origin, as well as a singular one
 $f_2 (r\rightarrow 0) \sim
c_1  + c_2 r^{-3}$. The long distance behaviour is $f_2 (r\rightarrow
\infty) \sim c_1{'} r^{G-1} e^{-Gr} + c_2{'} r^{-2} e^{-r}$.
The corresponding behaviour of $f_1$ is
$f_1 (r\rightarrow \infty)
\sim d_1 r^{G-1} e^{-(G+1)r} + d_2 r^{-2}$, where $d_1$ is determined in terms
of $c_1'$ and $d_2$ is determined in terms of $c_2'$.   The
$c_1'$ component comes from the exponentially increasing solution
of \schr, $u\sim e^{{1-G \over 2} r}$  and is non normalizable for
$G=0$. For non-zero $G$, both solutions for
$f_1$ and $f_2$ are  well behaved
at infinity, so in particular the solution regular at the origin is
normalizable. For $G=0$ only one solution is normalizable at infinity.
 Although $f_1$ is always
well behaved at long distances,  $f_2$ is only normalizable
for the choice $c_1' = 0$. Thus an  acceptable  solution to the system
can only exist if  the solution regular at the origin happens to be the
one regular at infinity. But if such a regular $f_2$ existed, the
corresponding $u(r)$ obtained from \udef\ would also be normalizable.
The behaviour of the effective potential in \schr\ rules this out.
Indeed, since $V_{\rm eff} (r) > (1-G)^2 /4$ we cannot have a zero
energy solution of \schr\ which is regular both at the origin and at
infinity. This means that, if $c_2=0$,
then  $c_1^{'} \neq 0$ in the $G\rightarrow 0$
limit and $f_2 (r)$ is not normalizable.

 As a consequence, it seems that  the only
normalizable fermionic zero modes in the single  monopole sector are those
generated by supersymmetry. This circumstance greatly simplifies the
quantization of the associated collective coordinates, since explicit
fermionic solutions of the equations of motion can be written down by
simple supersymmetry
rotations of the original bosonic solution. To second order in
the fermionic parameter $\epsilon$ we find
\eqn\fansatz
{\eqalign{A_{\epsilon}^i & = A^i_{PS} +
 {1\over 2M}\,  \epsilonb\, B^i_{PS}\,\epsilon \cr
 A_{\epsilon}^0  &= {1\over 2M}\, \epsilonb\, \sigma^i  B^i_{PS}
\,\epsilon\cr
\phi_{\epsilon} &= \phi_{PS} + {i\over 2M}\, \epsilonb\, \sigma^i  D_i
\phi_{PS}\,\epsilon  \cr
\tphi_{\epsilon} &= \tphi_{PS} + {i\over 2M}\, \epsilonb
\, \sigma^i  \tD_i
 \tphi_{PS}\,\epsilon }}
This solution, together with \zmodes, can be made into a collective
coordinate ansatz by simply giving time dependence to $\epsilon$.
Upon direct substitution into the field theory action we obtain the
following effective lagrangian for the fermionic coordinates:
\eqn\feff
{\CL_{\rm eff} = i\, \epsilonb\, \partial_t \,\epsilon }
where $\epsilonb\epsilon = \epsilonb_{\dot\alpha} \delta_{{\dot
\alpha} \alpha} \epsilon_{\alpha}$. Canonical quantization of this
fermionic quantum mechanics leads to
\eqn\can
{\{\epsilon_{\alpha}, \epsilonb_{\dot\alpha}\} = \delta_{\alpha{\dot
\alpha}} }
so that $\epsilonb_{\dot\alpha}$, ${\dot \alpha}= 1,2$ generates a
Fock space of four states. Since $\epsilon$ was the supersymmetry
parameter, if the vacuum is regarded as a scalar (corresponding to
the spherically symmetric PS solution), then we have two spin zero
 states
and two polarizations of spin $1/2$, a standard massive representation
of $N=1$ supersymmetry. The  spin operator
acts in the fermionic quantum mechanics as
$
J_3 = {1\over 2} \epsilonb \sigma_3 \epsilon
$,
and satisfies $[J_3, \epsilon_{\alpha}] = -{1\over 2} (\sigma_3
\epsilon)_{\alpha}$. Altogether, we find the degrees of freedom of a
massive chiral superfield.

The bosonic collective coordinates are easily introduced by giving an
implicit time dependence to all bosonic fields in terms of the
position collective coordinate $X^i (t)$. In practice, this amounts to
the following rule for time derivatives:
$$
\partial_t ({\rm Boson}) = {\dot X}^i \partial_i ({\rm Boson})
$$
We can also consider $U(1)$ charge rotations at infinity by an angle
$\chi \in [0,2\pi)$, leading to dyon solutions, which simply modify
the ansatz \fansatz\ by the addition of a term ${\dot \chi} \phi_{PS}$
to $A_0$.
 Finally, the ansatz for $A_0$ should be further corrected in
order to satisfy Gauss law, with the final result:
\eqn\acero
{A_0 (\epsilon,\chi, X^i)
 = -{1\over 2M}\, \epsilonb\, \sigma^i  B^i_{PS}\,\epsilon +
{\dot\chi}\phi_{PS} + {\dot X}^i A^i_{PS} }
where we have retained only terms of first order in time derivatives,
and second order in anticommuting parameters. This is enough to
satisfy the field equations to this order, and further ensure that the
collective motion is orthogonal to gauge transformations. Plugging the
complete ansatz into the field theory action, and integrating over
space we arrive at the complete effective quantum mechanics in the
one-monopole sector
\eqn\lefff
{\CL_{\rm eff} = {M\over 2} {\dot X}^i {\dot X}^i + {M\over 8}
{\dot\chi}^2 + i\,\epsilonb\, {\dot\epsilon} }
where we have used the BPS relations for the monopole mass:
    $M= \int
d^3 x (B_{PS})^2 = 4 \int d^3 x (D\phi_{PS})^2 $.
 After quantization, the conjugate of $X^i$ becomes the standard space
momentum, while the conjugate of $\chi$, a compact coordinate, becomes
the quantized electric charge of the dyon.

\newsec{Global quantum numbers}

The effective fermionic quantum mechanics is useful to calculate low
energy dynamics of monopoles and dyons. Even in the single monopole
sector it gives important information, like the quantum numbers of the
soliton with respect to the unbroken global symmetries. In this case
the unbroken flavour group $SU(N_f -1)_L \times SU(N_f -1)_R$ and two
$U(1)$ symmetries: the unbroken baryon number and R-symmetry. We
simply need to represent the corresponding charges in the quantum
mechanical Fock space. Following ref. \refs\rjackreb, this is easily
accomplished by direct reduction of the field theory Noether currents
in the collective coordinate ansatz. For example, the unbroken flavour
currents in the field theory are
$$
j_R^{a\mu} = {1\over 2} \sum_{r,s=2}^{N_f} \left( \psib^r (t^a)_{r}^s
\sigmab^{\mu} \psi_s - \psi_r (t^a)_{r}^s \sigma^{\mu} \psib^r \right)
+ {\rm bosons}
$$
$$
j_L^{a\mu} = {1\over 2} \sum_{r,s=2}^{N_f} \left( \tpsib_s
(t^a)_{r}^s \sigmab^{\mu} \tpsi^r - \tpsi^r (t^a)_{r}^s \sigma^{\mu}
\tpsib_s \right) + {\rm bosons}
$$
Since the fermionic zero modes have no component along  the
$r=2,3,...,N_f$ flavours, the currents act trivially on the monopole
chiral superfield: it is a singlet of $SU(N_f -1)_L \times SU(N_f
-1)_R$.
In a similar fashion, the general $U(1)$ current is given by
\eqn\uuno
{\eqalign{ j^{\mu} &= {q_{\lambda}\over 2} \left( \lambdab^a
\sigmab^{\mu} \lambda^a - \lambda^a \sigma^{\mu}\lambdab^a \right) +
\sum_{r=1}^{N_f} {q_{\psi}^r\over 2} \left( \psib^r \sigmab^{\mu}
\psi_r - \psi_r \sigma^{\mu}\psib^r\right) +{ {\tilde q}_{\psi}^r
\over 2}\left(\tpsib_r \sigmab^{\mu}\tpsi^r - \tpsi^r \sigma^{\mu}\tpsib_r
\right) \cr
&-iq_{\phi}^r \left( ( D^{\mu} \phi^{\dagger})^r  \phi_r -
\phi^{\dagger r} D^{\mu} \phi_r \right) -i {\tilde q}_{\phi}^r
\left( \tphi^r \tD^{\mu} \tphi_r^{\dagger} - (\tD^{\mu}\tphi)^r
\tphi_r^{\dagger} \right) }}
where all the currents are defined as normal ordered with respect to the
perturbative vacuum. The original baryon number symmetry with charges
$q_{\lambda} =0$, $q_{\psi}^r = q_{\phi}^r = -{\tilde q}_{\psi}^r
=-{\tilde q}_{\phi}^r = 1$ is broken in the monopole configuration
by the squark spectation values $\bra \phi_1 \ket = \bra \tphi_1 \ket
\neq 0$. One can  define an unbroken $U(1)_{B'}$ by combining
the previous one with a vector $SU(N_f)$ transformation such that
$\phi_1$ and $\tphi_1$ remain invariant. The corresponding charge
asignments are $(q')^1 =- ({\tilde q}')^1 = 0$ and
 $q' =- {\tilde q}'
= {N_f \over N_f -1} $ for the rest of the flavours. Again, since the
supersymmetry zero modes are non zero only along the $r=1$ flavour
component, the new currents act trivially and the monopole has baryon
number zero. The situation is different for the  R-symmetry. The
original anomaly free R-symmetry has charges
$$
q_{\lambda} =1 \,\,\,, \,\, q_{\phi}^r = {\tilde q}_{\phi}^r =
{N_f - N_c \over N_f}
$$
 with the usual grading $q_{\psi} = q_{\phi} -1$ within the
chiral supermultiplet. The $r=1$ squark expectation values break this
symmetry, which nevertheless can be combined with an axial $SU(N_f)$
trasformation to define an unbroken $U(1)_{R'}$ with charges
$q'_{\lambda} =1$, $ (q')^1 = ({\tilde q}')^1 = 0$ and $q' =
{\tilde q}' = {N_f -N_c \over N_f -1}$ for the rest of the
flavours. Note that this
 $R'$ symmetry is non-anomalous with the new massless field
content left after Higgs mechanism. The charge operator acting on the
fermionic Fock space can be readily calculated as
\eqn\rcharge
{Q_{R'} = {1\over 2} \int d^3x \left( \lambdab_{\epsilon}^a
 \lambda_{\epsilon}^a - \lambda_{\epsilon}^a
\lambdab_{\epsilon}^a \right)- {1\over 2}\int d^3x
\left(\psib_{\epsilon} \psi_{\epsilon} -
\psi_{\epsilon} \psib_{\epsilon} + \tpsib_{\epsilon} \tpsi_{\epsilon}
- \tpsi_{\epsilon}\tpsib_{\epsilon}\right) = \epsilonb\, \epsilon -1}

The R charge of the monopole vacuum is $-1$, and the grading within the
massive supermultiplet is given by
$$
[Q_{R'} , \epsilon_{\alpha}] = -\epsilon_{\alpha}
\,\,\,,\,\,\,\,[Q_{R'}, \epsilonb_{\dot\alpha}] =
\epsilonb_{\dot\alpha}
$$
We conclude that the monopole chiral superfield has R charge one,
compatible with a usual  mass term in a dual effective lagrangian
description of the form $W_m \sim m Y^2$.

\newsec{Discussion}
At present, we do not have a concrete proposal regarding the
role of these solutions in the general picture of duality.
We make some brief comments on the possibility
of seeing semiclassical signals of $N=1$ duality in the sense
of finding states which resemble the dual  magnetic degrees
of freedom.
 The semiclassical analysis of monopoles
looks at an object whose mass increases at large Higgs
vev, and for asymptotically free theories the semiclassical
methods become more reliable for large vevs. On the other hand
$N=1$ duality is a statement about the theory in the far infrared.
So there are two possible scenarios in which one can
compare   the semiclassical
calculations to duality. One is to compare with possible extensions
of duality beyond the far  infrared, and the other is to consider
the behaviour  of the states constructed in the semiclassical
quantisation as the Higgs vev is tuned to zero, and we flow towards the
origin of moduli space. Because of
the lack of stability and BPS saturation property
 the semiclassical monopoles are
not guaranteed to define states which survive
in the strong coupling region. Since it is not impossible that
 they do survive and become stable at strong coupling by a currently
unknown mechanism,
we compare the monopoles constructed here with
some of the characters in $N=1$ duality.

 The simplest  class
of models to consider is $SU(N_c)$ with $N_f$ flavours, with
$N_c+2 \le N_f \le 3N_c$.
 The spin degrees of
freedom and the unbroken $U(1)_{R'}$ charge
are the same as those of the dual quark which acquires a mass
when the electric squark acquires an expectation value. But
the  baryon number of the heavy dual quarks under
$U(1)_{B'}$ is given by ${N_f \over N_f - N_c}$, while we found $B' =
0$ for the monopole. Another difference is that these quarks
come in quark-antiqark pairs related by charge conjugation, whereas the
monopole is self conjugate. Solutions with positive or negative
magnetic charge are gauge equivalent.

These monopoles can be embedded in the theories considered in
\refs\rkut , in which the electric theory has an adjoint $X$
and the magnetic theory has an adjoint $Y$, in addition to sets of
quarks and antiquarks.
The steps in the semiclassical treatment are very similar.
The
monopoles embedded in the electric theory have many of the properties
of the $Y$ field. The
$Y$ are adjoint chiral superfields, have baryon number zero
and are singlets under
the flavour group. However two problems remain in identifying the monopoles as
 candidates for $Y$. The R-charge of $Y$ under the
 unbroken $U(1)_{R'}$    is not $1$.  In characterizing
the gauge quantum numbers of these monopoles,  the dual  groups
of $SU(N)$ groups have the form $SU(N)/Z_N$, which are different
from the dual groups entering $N=1$ duality of \rseidos\ or \rkut\
even at the self-dual points
 ( for more details on this issue see section 3).

In summary, the physical interpretation of these solutions
is not straightforward. However a simple  characterization
of their transformation properties under Lorentz, global symmetry and
appropriate dual gauge groups is possible. It
would therefore be interesting to search for
 consistent scenarios relating  these and other
monopoles in $N=1$ theories to duality, perhaps by   embedding them
in theories which allow stability and yet are related
to super-QCD by some simple perturbation. It is also possible that the
unstable monopoles described here could find some application in
other contexts. For example, unstable monopoles in QCD were
proposed in \rgp\ as  a heuristic mechanism for the generation
of ``magnetic mass" in high temperature QCD.

\newsec{Acknowledgements}
We would like to thank P. van Baal,
O. Ganor, D. Gross, A. Losev, G. Moore,
D. Olive  for discussions.
This work was supported by NSF grant PHY90-21984  and DOE grant
DE-FG02-91ER40671.
\listrefs
\bye